\documentclass[onefignum,onetabnum]{siamart220329}

\usepackage{amsfonts}
\usepackage{graphicx}
\usepackage{algorithmic}
\usepackage{textcomp}
\ifpdf
  \DeclareGraphicsExtensions{.eps,.pdf,.png,.jpg}
\else
  \DeclareGraphicsExtensions{.eps}
\fi

\newcommand{\N}{\mathbb{N}}
\newcommand{\Z}{\mathbb{Z}}
\newcommand{\lb}{\mathrm{lb }}
\newcommand{\li}{\mathrm{li }}
\newcommand{\Li}{\mathrm{Li }}

\newsiamremark{remark}{Remark}
\newsiamremark{hypothesis}{Hypothesis}
\crefname{hypothesis}{Hypothesis}{Hypotheses}
\newsiamthm{claim}{Claim}

\headers{Optimizing the size of array for modern DFT libs}{A.\,O.~Korotkevich}

\title{Optimizing the size of array for modern discrete Fourier transform libraries\thanks{Submitted to the editors 2023-08-10.
\funding{This work was funded by Simons' Foundation through Simons' Collaboration on Wave Turbulence (award \#651459).}}}

\author{Alexander O. Korotkevich\thanks{Department of Mathematics and Statistics, University of New Mexico, Albuquerque, NM, USA 
  (\email{alexkor@math.unm.edu}); L.\,D.~Landau Institute for Theoretical Physics, Russian Academy of Sciences, Chernogolovka, Moscow region, Russia.}}

\ifpdf
\hypersetup{
  pdftitle={An Example Article},
  pdfauthor={D. Doe, P. T. Frank, and J. E. Smith}
}
\fi

\begin{document}

\maketitle

\begin{abstract}
The problem of optimization of the array size for modern discrete Fourier transform libraries is considered and reformulated as an integer linear programming problem. Acceleration of finding an optimal solution using standard freely available library with respect to brute force approach is demonstrated. Ad hoc recursive algorithm of finding the optimal solution is proposed, complexity scaling of the algorithm is estimated analytically. The problem can be used in a linear programming class as an example of purely integer programming problem (continuous linear programming solution has no sense), simple enough to be solved using even interpreting programming languages like Python or Matlab.
\end{abstract}




\section{Introduction\label{sec:intro}}
Modern discrete Fourier transform (DFT) libraries provide $\mathcal{O}(N\log N)$ complexity scaling for any integer size of the transform $N$. At the same time the coefficient in front of the $N\log N$ expression can differ strikingly (see, for example, results for 2D DFTs with sizes $(1024\times K)^2$, where $K\in \N$, Figs. 4-6 in~\cite{2DFourier2020}) depending whether the number $N$ can be represented as a product of integer (non negative) powers of the first prime numbers. Specifically, this is a quote~\cite{FFTW_primes} from the documentation of the FFTW~\cite{FFTW}, which is the de facto standard in pseudo spectral~\cite{Boyd} methods:\\
{\sl ``The input data can have arbitrary length. FFTW employs $\mathcal{O}(n \log n)$ algorithms for all lengths, including prime numbers.''\\
``For example, the standard FFTW distribution works most efficiently for arrays whose size can be factored into small primes (2, 3, 5, and 7), and otherwise it uses a slower general-purpose routine. If you need efficient transforms of other sizes, you can use FFTW’s code generator, which produces fast C programs (“codelets”) for any particular array size you may care about. For example, if you need transforms of size $513 = 19\times3^{3}$, you can customize FFTW to support the factor 19 efficiently.''}

This is a practical problem similar to which author encountered in research for~\cite{2DFourier2020,KLSD2023}.
Let us consider a computational workstation with 10 TB of RAM. Let us say that our computational code requires three 2D arrays of data for some (pseudo-) spectral code. What is the maximum size of array which we can use for computation?
\begin{itemize}
\item
$10$ TB = $10\times 2^{40}$ Bytes.
\item
Let us consider complex-to-complex Fourier transforms: every complex number is represented by two floating point numbers of double precision. IEEE 754 floating point number of double precision occupies 8 Bytes. So complex number of double precision occupies $16=2^{4}$ Bytes.
\item
$10\times 2^{40}/2^{4} = 10\times 2^{36}$ complex numbers of double precision can fit into memory.
\item
Let us consider, for the sake of simplicity, square 2D arrays (recall that we need three of them).
\item
$\sqrt{10\times 2^{36}/3}\simeq 478607.27$
\end{itemize}
The same logic can be applied to 3D (or n-dimensional) arrays as well. Multidimensional DFTs are computed using 1D DFTs, so the optimal size problem stays the same.

Taking into account what we already know about efficiency of FFTW (and other DFT libraries), to maximize performance of the DFTs the previous problem can be reformulated as follows:
\begin{itemize}
\item
Find $N\in \N$ closest to $478607.27$ subject to the following conditions:
\item
$N\le 478607.27$ (we still need to fit into memory)
\item
$N=2^{x_2}3^{x_3}5^{x_5}7^{x_7}$, where $x_i \in \Z_{+}=\N\cup\{0\}=\{0,1,2,\ldots\}$, for $\forall i\in\{2,3,5,7\}$. 
\end{itemize}
Taking into account the mentioned above fact that one can ``teach'' FFTW to use more primes efficiently, we need to be able to generalize the problem accordingly. In this paper we will consider such generalized problem; reformulate it as a canonical linear program for discrete unknowns belonging to $\Z_{+}$; prove existence and uniqueness of the optimal solution; propose ad hoc recursive algorithm, which appears to be more efficient than generic algorithm for 
solution of integer linear programming problems; evaluate computational complexity of the proposed algorithm.

The paper is organized as follows. Different formulations of the problem are in~\cref{sec:formulation}, our new algorithm is described in \cref{sec:alg} (its computational complexity is evaluated in details in~\cref{apdx:complexity}), numerical
results are in~\cref{sec:experiments}, and the conclusions follow in~\cref{sec:conclusions}.

\section{Different formulations of a problem}
\label{sec:formulation}

Let us consider a set of $n$ distinct prime numbers ${p_1, p_2, ..., p_n}$ and a real number $D\in \mathbb{R}$ such that $\exists i\in [1,\ldots,n]:\, p_i\le D$.
We need to find maximum $N=p_1^{x_1}\times p_2^{x_2}\times\cdots\times p_n^{x_n}$, where $\forall i\in [1,\ldots,n],\, x_i\in\Z_{+}$, subject to condition $N\le D$. From the last condition one could immediately derive constraints for powers of primes: $0 \le x_i \le \lfloor \log_{p_i} D \rfloor$, where $\lfloor\cdot\rfloor$ denote a standard floor or rounding down to the nearest integer function. As a result one can state the following {\bf nonlinear} program:
\begin{equation}
\label{NLPformulation}
\begin{array}{ll@{}ll}
\text{maximize}  & \displaystyle p_1^{x_1}\times p_2^{x_2}\times\cdots\times p_n^{x_n} &\\
\text{subject to}& \displaystyle p_1^{x_1}\times p_2^{x_2}\times\cdots\times p_n^{x_n} \le D &,\\
                 & \displaystyle 0 \le x_i \le \lfloor \log_{p_i} D \rfloor,  &i=1 ,\ldots, n\\
                 &                                                x_{i} \in \Z_{+}, &i=1 ,\ldots, n.
\end{array}
\end{equation}
Upper bound for $x_i$ is redundant and follows directly from the main constraint $N\le D$, but helpful for computational complexity analysis, so we will keep it from now on. Total number of sets of prime powers to be checked for optimality is
\begin{equation}
\label{BF_complexity}
C_{BF} = \prod\limits_{i=1}^{n}(\lfloor \log_{p_i} D \rfloor + 1).
\end{equation}
This problem can be solved directly by the brute force (this why $C_{BF}$ for complexity) check of every set of $x_i$'s. Using more optimal approaches for computing integer powers of primes (also integers) like exponentiation by squaring~\cite{Knuth_v2} instead of general power functions can provide a significant speedup.

One can transform the problem to the linear one by applying logarithm function to both sides of functional and the first constraint (because $N\in\N$ and logarithm is a monotonous function resulting inequality will hold). Any base of the logarithm larger than  can be used, but for simplicity let us use $\log_2 = \lb$ (here and further we will follow DIN 1302, ISO 31-11, and ISO 80000-2 standards for notation of binary logarithm). Binary logarithm can be computed somewhat more efficiently and is one of the standard functions in most of popular programming languages. Also, in all DFT realizations author had experience with at least one of efficient primes is $2$, which allows us to simplify the formulae. In some of the early realizations this is the only efficient prime $p_i$. The resulting {\bf linear} program has the following form:
\begin{equation}
\label{LPformulation}
\begin{array}{ll@{}ll}
\text{maximize}  & \displaystyle x_1\lb p_1+ x_2\lb p_2 +\cdots + x_n\lb p_n &\\
\text{subject to}& \displaystyle x_1\lb p_1+ x_2\lb p_2 +\cdots + x_n\lb p_n \le \lb D &,\\
                 & \displaystyle 0 \le x_i \le \left\lfloor \frac{\lb D}{\lb p_i} \right\rfloor,  &i=1 ,\ldots, n\\
                 &                                                x_{i} \in \Z_{+}, &i=1 ,\ldots, n.
\end{array}
\end{equation}
For simplicity we represented upper bound for $x_i$ through binary logarithm. Now ``computational complexity'' (total number of sets of prime powers to check for optimality) can be expressed as follows:
\begin{equation}
\label{BF_complexity_lb}
C_{BF} = \prod\limits_{i=1}^{n}\left(\left\lfloor \frac{\lb D}{\lb p_i} \right\rfloor + 1\right).
\end{equation}
It should be noted, that in this formulation one needs only multiply and subtract real numbers in order to find a mismatch with respect to $D$, so in reality, although numbers of optimality checks is the same, such computation will be faster. 

This is a canonical form of a linear program~(see, e.g. \cite{KarloffLP}). It is worth noting, that the problem has no unique solution if we consider it as an ordinary linear program with continuous $x_i\in\mathbb{R}$. One can put all but one unknown $x_\alpha$ to zero and then put $x_\alpha = \log_{p_\alpha} D$. Let us show existence and uniqueness of the optimal solution of the integer linear programming problem~\cref{LPformulation}. Existence follows from the condition that at least one of $p_i$'s is less or equal to $D$. Let us say it is true for at least $p_{\gamma}$. It means that we have a basic solution $x_i = 0$ for $i\ne\gamma$ and $x_\gamma = \lfloor\log_{p_\gamma} D\rfloor$, which can be optimized going through all possible $0 \le x_i \le \lfloor \log_{p_i} D \rfloor,\,  i=1 ,\ldots, n$. Uniqueness can be proven in the same way as uniqueness of factorization of an integer number into primes (Fundamental Theorem of Arithmetic, also known as the Unique Factorization Theorem or Prime Factorization Theorem). Let us consider two representations $\vec x = [x_1,\ldots,x_n]$ and $\vec y = [y_1,\ldots,y_n]$ of the optimal solution $N$:
\begin{align*}
N = p_1^{x_1}\times p_2^{x_2}\times\cdots\times p_n^{x_n} = p_1^{y_1}\times p_2^{y_2}\times\cdots\times p_n^{y_n},\\
1 = p_1^{x_1-y_1}\times p_2^{x_2-y_2}\times\cdots\times p_n^{x_n-y_n}.
\end{align*}
The last expression can be valid only in the case $x_i=y_i, \forall i=1 ,\ldots, n$, because otherwise we have either irreducible fraction or some positive power of one or several of the primes $p_i$'s to be equal to $1$.
This unique solution can be found by any generic integer linear programming algorithms, for example utilizing Gomory's cuts~(see, e.g. \cite{WolseyIP}).

\section{Ad Hoc Algorithm}
\label{sec:alg}
Let us try to improve (decrease) the number of operations for finding an optimal solution of~\cref{LPformulation}. For example, let us suppose, the first $(n-1)$ primes' powers $x_i,\, i=1,\ldots,n-1$ are set to some values (and fixed). In such a case an optimal value of the last power $x_n$ can be calculated by the following formula:
\begin{align}
x_n &= \left\lfloor\log_{p_n}\left(\frac{D}{p_1^{x_1}\times p_2^{x_2}\times\cdots\times p_{n-1}^{x_{n-1}}}\right)\right\rfloor=\nonumber\\
&=\left\lfloor\log_{p_n}D - x_1\log_{p_n}p_1 - x_2\log_{p_n}p_2 - \ldots - x_{n-1}\log_{p_n}p_{n-1}\right\rfloor=\nonumber\\
&=\left\lfloor\frac{1}{\lb p_n}(\lb D - x_1\lb p_1 - x_2\lb p_2 - \ldots - x_{n-1}\lb p_{n-1})\right\rfloor.\label{x_n_optimal}
\end{align}

Now, let us consider a little bit more complex case, when only the first $(n-2)$ primes' powers are set to some values $x_i,\, i=1,\ldots,n-2$. In such a case unknown $x_{n-1}$ can take values from $0$ to
\begin{align}
x_{n-1}^{\max}&=\left\lfloor\log_{p_{n-1}}\left(\frac{D}{p_1^{x_1}\times\cdots\times p_{n-2}^{x_{n-2}}}\right)\right\rfloor=\nonumber\\
&=\lfloor\log_{p_{n-1}}D - x_1\log_{p_{n-1}}p_1 - \ldots - x_{n-2}\log_{p_{n-1}}p_{n-2}\rfloor=\nonumber\\
&=\left\lfloor\frac{1}{\lb p_{n-1}}(\lb D - x_1\lb p_1 - \ldots - x_{n-2}\lb p_{n-2})\right\rfloor.\label{x_n_1_upper_bound}
\end{align}
For each value of $x_{n-1}$ we need to use~\cref{x_n_optimal} in order to determine optimal value of $x_n$ and find mismatch with $D$. Optimal choice of $x_{n-1}$ will have a minimal mismatch (equivalent of a condition $\max(N)$ with a constraint $N\le D$). One can note that if we decide to use binary logarithm, then we can compute values $\lb D, \lb p_1, \ldots,\lb p_n$ once a use them in all computations. Due to monotonicity of a logarithm function, as an equivalent of a mismatch (or residue) $\Delta N_n$ for minimization we can use the following expression:
\begin{equation}
\Delta \tilde N_n = \lb D - x_1\lb p_1 - x_2\lb p_2 -\ldots -x_n\lb p_n.\label{logarithm_of_mimatch}.
\end{equation}

The number of evaluations of a mismatch (or checks for optimality of a solution) for the case of the fixed first $(n-2)$ primes' powers is
$$
\sum\limits_{x_{n-1}=0}^{x_{n-1}^{\max}} 1 = x_{n-1}^{\max} + 1.
$$

It is convenient to introduce a value $a_k$:
\begin{equation}
a_k = \lb D - x_1\lb p_1 - x_2\lb p_2 -\ldots -x_{k-1}\lb p_{k-1},\label{a_k_definition}
\end{equation}
so, e.g. $x_{n-1}^{\max} = \lfloor a_{n-1}/\lb p_{n-1}\rfloor$. If the first $m$ primes' powers are fixed, the $x_{m+1}$ has to take values from $0$ to $x_{m+1}^{\max}=\lfloor a_{m+1}/\lb p_{m+1}\rfloor$, for each value of $x_{m+1}$ we need to find optimal set of the remaining powers $x_{m+2},\ldots,x_{n}$. This brings us to the idea of the following recursive function~\cref{alg:recursive}, which for for fixed first $m$ primes' powers finds optimal values of $x_{m+1},\ldots,x_{n}$. As an input parameters we will use $\Delta \tilde N_{m}$ (see~\cref{logarithm_of_mimatch}), number of the first fixed primes' powers $m$, and the solution vector $\vec x = [x_1,\ldots,x_n]$, containing the first $m$-powers (which are considered given and will not be changed) and anything in remaining components; the function will set optimal values of $x_{m+1},\ldots,x_{n}$ for given $x_1,\ldots,x_{m}$, thus giving us basic solution (which might be the optimal one) $\vec x = [x_1,\ldots,x_n]$, where the $x_1,\ldots,x_m$ might be suboptimal, but the $x_{m+1},\ldots,x_{n}$ will be the optimal choice of remaining unknowns. 
\begin{algorithm}
\caption{Ad hoc recursive algorithm {\tt optimal\_powers ($\Delta \tilde N_{m}$, $m$, $\vec x$)}}
\label{alg:recursive}
\begin{algorithmic}
\IF {$m = n-1$}
  \STATE {set $n$-th component in $\vec x$ to $x_n$ using~\cref{x_n_optimal} and exit.}
\ELSE
  \STATE {$x_{m+1}^{\max} := \lfloor \Delta \tilde N_{m}/\lb p_{m+1} \rfloor$;}
  \STATE {$x_{m+1}:=0$;}
  \STATE {$\Delta \tilde N_{\min} := \Delta \tilde N_{m}$; /* setting current minimal mismatch */}
  \STATE {$\vec x_{\mathrm{optimal}}:=\vec x$;}
  \WHILE {$x_{m+1}\le x_{m+1}^{\max}$}
    \STATE {set $m+1$-th component in $\vec x$ to $x_{m+1}$;}
    \STATE {recursive call {\tt optimal\_powers ($\Delta \tilde N_{m} - x_{m+1}\lb p_{m+1}$, $m+1$, $\vec x$)};}
    \IF {mismatch $\Delta \tilde N_{n}$ computed using components of returned $\vec x$ by~\cref{logarithm_of_mimatch} $\le \Delta \tilde N_{\min}$}
      \STATE {$\vec x_{\mathrm{optimal}}:=\vec x$;}
      \STATE {$\Delta \tilde N_{\min} := \Delta \tilde N_{n}$;}
    \ENDIF
    \STATE {$x_{m+1}:= x_{m+1} + 1$;}
  \ENDWHILE
  \STATE {$\vec x := \vec x_{\mathrm{optimal}}$;}
\ENDIF
\end{algorithmic}
\end{algorithm}
Now in order to find optimal set of all we need to do is to call a function {\tt optimal\_powers} with the following parameters:\\
{\tt optimal\_powers ($\lb D$, $0$, $\vec x$)}\\
which will result in optimal set of powers of primes stored in $\vec x$.

Number of evaluations (checks for optimality) of the ad hoc recursive algorithm will be the following:
\begin{align}
C_{AHR}&=\sum\limits_{x_1=0}^{\lfloor a_{1}/\lb p_{1}\rfloor}\sum\limits_{x_2=0}^{\lfloor a_{2}/\lb p_{2}\rfloor}\cdots\sum\limits_{x_{n-1}=0}^{\lfloor a_{n-1}/\lb p_{n-1}\rfloor} 1\approx\nonumber\\
&\approx \frac{1}{(n-1)!}\prod\limits_{i=1}^{n-1}\frac{\lb D}{\lb p_i} = \frac{1}{(n-1)!}\prod\limits_{i=1}^{n-1}\log_{p_i} D,\label{AHR_complexity}
\end{align}
here $\approx$ sign means that we limit ourselves by the leading term. Complete derivation can be found in~\cref{apdx:complexity} Comparing~\cref{AHR_complexity} and~\cref{BF_complexity_lb} we can evaluate a speedup factor:
\begin{equation}
\label{AHR_speedup}
\frac{C_{BF}}{C_{AHR}} \approx (n-1)!\frac{\lb D}{\lb p_n} = (n-1)!\log_{p_n} D.
\end{equation}

\section{Experimental results}
\label{sec:experiments}

In order to compare performance of brute force (BF) algorithms, proposed recursive ad hoc algorithm, and generic integer linear programming approaches, the code in C programming language was written. The choice of the programming language was due to used implementation of a generic solver for integer linear programming problems from the GNU Linear Programming Kit (GLPK)~\cite{GLPK}, which is written in C as well. The description of how to compile and use the code, which is included as a supplemental material for the paper, is provided in~\cref{apdx:code}.

In order to decrease the influence of a system processes and possible CPU frequency changes due to throttling, ain every run every method was executed  1000 times, the average value reported (total time for a run of every method divided by 1000), procedure repeated 3 times one after another and the best values taken as the final results. All computations were performed on a 12th Generation Intel\textregistered{} Core\texttrademark{} i5-12600H (memory usage was low, so 16GB RAM is not a relevant parameter), which is a mobile CPU.

Results of application of different algorithms to the problem formulated in Introduction~\cref{sec:intro} are shown in~\cref{tab:numerics}.
Obviously, all algorithms delivered the same solution: $476280=2^{3}\times 3^{5}\times 5^{1}\times 7^{2}\times 11^{0}$.
First of all, it is interesting to note that although generic {\tt pow()} function in C is well optimized, because it uses floating point arithmetics, it is slower that even relatively simple implementation of computation of an integer power of an integer (in current implementation {\tt unsigned long int} data type) number using exponentiate by squaring algorithm~\cite{Knuth_v2}. Speedup for brute force solution of~\cref{NLPformulation} is approximately 24\%. More optimized version of the function can slightly improve the performance. Secondly, just a simple replacement of a problem formulation by a linear version~\cref{LPformulation} (so instead of multiplication of powers of primes one computes sum of products of integer powers and precomputed values of binary logarithms of primes) gives speedup 58\% without writing new functions! Thirdly, generic solver for integer linear programming problem from GLPK is almost four (3.9) times faster than the fastest brute force approach, which uses linear programming formulation.

\begin{table}[htbp]
\footnotesize
\begin{center}
  \begin{tabular}{|c|l|c|}
  \hline
   & Algorithm & time, s \\
   \hline
    1 & BF, generic pow-function & $6.24\times 10^{-03}$ \\
    2 & BF, simple integer pow-function & $4.75\times 10^{-03}$ \\
    3 & BF, optimized integer pow-function & $4.70\times 10^{-03}$ \\
    4 & BF, LP formulation through $\lb$ & $2.63\times 10^{-03}$ \\ 
    5 & AHR, increasing primes & $1.25\times 10^{-05}$ \\
    6 & AHR, decreasing primes & $2.91\times 10^{-06}$ \\
    7 & GLPK, integer linear programming & $6.71\times 10^{-04}$ \\
    \hline
  \end{tabular}
\end{center}
\caption{Time in seconds (rounded to two digits after decimal point) for finding the unique solution of a problem~\cref{LPformulation} using different computational approaches. BF stands for brute force, AHR means recursive ad hoc algorithm. $D=478607.27$ (see~\cref{sec:intro} for explanation), set of primes was $p_1=2, p_2=3, p_3=5, p_4=7, p_5=11$ for the case of increasing and $p_1=11, p_2=7, p_3=5, p_4=3, p_5=2$ for the case of decreasing primes.}\label{tab:numerics}
\end{table}

Ad hoc recursive algorithm several orders of magnitude faster than even the fastest version of brute force approach. What is also interesting, it is sensitive to the ordering of primes (all other approaches are not, which is natural at least for brute force algorithms). For example, one can derive this fact from the estimated speedup with respect to brute force~\cref{AHR_speedup} (e.g. $p_n$ may be 2 or 11 in our case, depending upon what is the ordering). Another way to understand it follows from the estimation of the number of values for $x_1$ we need to go through (pay attention, it is independent from values of other powers), which is $(\lfloor\log_{p_1} D\rfloor + 1)$. Different $p_1$'s will give different length of the loop on $x_1$ (for large D the factor is approximately the same as using~\cref{AHR_speedup}, namely $\log_2 D/\log_{11} D = \log 11/\log 2\approx 3.46$). Both these estimations are rather crude, the real speedup is higher, approximately $4.3$. It should be noted, that the rough estimation~\cref{AHR_speedup} underestimates the real speedup of an ad hoc recursive algorithm approximately by a factor close to $2$. The best way of primes ordering is to have them monotonously decreasing, in such a case we have almost three orders (around $904$ times, while the formula~\cref{AHR_speedup} gives the value of $\approx 453$) of magnitude speedup with respect to the fastest brute force approach (which is through linear programming formulation~\cref{LPformulation}). If we order primes oppositely (monotonously increasing), corresponding speedup with respect to the fastest brute force implementation is ``only'' a little bit more that two orders of magnitude (around $210$ times, compare with estimation $\approx 131$ using~\cref{AHR_speedup}).

Still for the worst brute force approach the time of single problem solution is very small, so even algorithms implemented using interpreting programming languages supposed to be not too slow, which makes the problem accessible during a class on (integer) linear programming of undergraduate level.

\section{\label{sec:factorization} On application of the proposed ad hoc recursive algorithm to primes decomposition}
In modern cryptography the primes factorization or primes decomposition of a positive integer number into product of powers of primes is a very important problem~\cite{IntFactorization}, which is used in RSA algorithms for public key encryption and digital signature. It should be noted that the problem which we discuss in this paper is a different one. We are trying to find closest factorization into {\bf given set of small primes}, while in cryptography product of few large primes is usually used as a key (product of two large primes is considered to be the hardest to factorize) and these primes are unknown. While our proposed ad hoc recursive algorithm has complexity which is going down (with respect to brute force algorithm) very fast with the number of used primes (as $1/(n-1)!$, where $n$ is the number of primes), the number of all previous primes for a large number (long key) is huge and will require immense storage. Even more, to find all of them requires a lot of computational resources. The number of primes smaller than some large number can be estimated using the Prime Number Theorem, which gives several asymptotics for $\pi(N)$ -- a prime counting function, which gives
number of primes less than $N$:
$$
\pi(N) \sim \frac{N}{\ln(N)}\sim \Li(N)=\li(N)-\li(2),
$$
here $\ln$ is the natural logarithm (again, we use notation $\ln$ for a natural logarithm recommended in ISO 80000-2 and other standards), $\Li$ is the offset logarithmic integral expressed through logarithmic integral $\li$, also know as integral logarithm function~\cite{AbramowitzStegun}. $\Li(N)$ estimation converges way faster and although underestimates the $\pi(N)$ is way closer to the exact value.

So the proposed algorithm hardly can be useful for primes decomposition, unless one needs to find closest number which can be represented as product of some set of large precomputed primes.

\section{Conclusions}
\label{sec:conclusions}
We have considered a practical optimization problem, which was formulated as a nonlinear and then as a linear programming problem. It was shown that the problem has no sense if formulated for a continuous unknowns. Vice versa, if formulated properly, i.e. on a set of integers, the existence and uniqueness of solution was demonstrated. Both these facts make the problem potentially useful for classes on (integer) linear programming. Firstly, unlike the most of the model integer linear programming problems in textbooks (which historically originate from business tasks), the motivation comes from DFT libraries, which are used for signal processing, computational physics and optics, and computer science applications. As usual for similar problems of number theory, it is easy to formulate and comprehend, while it is hard or impossible to get analytical results.

It was demonstrated that even for brute force algorithms one can get significant speedup (from 24\% to almost 60\%) with respect to the most straightforward approach just investing a little bit of time into either optimizing the obvious bottlenecks (power function) via some classic seminumerical algorithms (two versions are given in the code, but there are way more optimized implementations, for example~\cite{OptimizedIntPow} and figuring out why it is written like that can be a topic for a whole separate lecture on code optimization, including considering approaches like unrolling loops, bit operations, switch fall through, visibility scope and static arrays etc.) or reformulating in a style of linear programming. Such discussions could be interesting and stimulating for students.

Generic integer linear programming solvers (implementing some flavour of Gomory cuts~\cite{WolseyIP}) give several times speedup with respect even to the fastest formulation of a brute force algorithm for the given size of the problem and number of primes. It is instructive to show that if one decrease the parameter $D$ (``size'' of the problem) or number of primes, one can get a situation when overhead for setting up the problem for a generic integer programming solver can eliminate all the speedup or make brute force approaches even preferable.

From the author's point of view, the most interesting part is the demonstration of the fact that relatively simple ad hoc algorithm, specialized for the problem, can be a lot faster than generic optimization subroutines from well established libraries. At the same time, such algorithms are usually working only for specific problem, while generic solver can be used for any problem in a wide class. 

Also, the proposed problem can be a reason to mention some classical problems of pure mathematics, specifically primes decomposition and number theory, which relatively recently become extremely applied in cryptography. This could bring a topic of the Prime Numbers Theorem, as very briefly discussed in~\cref{sec:factorization}.

\appendix
\section{\label{apdx:complexity}Derivation of complexity of an ad hoc recursive algorithm}
We would like to find approximation of an expression:
\begin{equation}
C_{AHR}=\sum\limits_{x_1=0}^{\lfloor a_{1}/\lb p_{1}\rfloor}\sum\limits_{x_2=0}^{\lfloor a_{2}/\lb p_{2}\rfloor}\cdots\sum\limits_{x_{n-1}=0}^{\lfloor a_{n-1}/\lb p_{n-1}\rfloor} 1.\label{apdx:start}
\end{equation}

The simplest and crudest way would be replace all summations by integrals and to drop $\lfloor\cdot\rfloor$ signs. This would give for the right most sum:
$$
\int\limits_{0}^{a_{n-1}/\lb p_{n-1}} 1 d x_{n-1} = \frac{a_{n-1}}{\lb p_{n-1}}.
$$
Taking into account, that according to~\cref{a_k_definition} $a_{n-1} = a_{n-2} - x_{n-2}\lb p_{n-2}$, the second from the right sum can be approximated as:
\begin{align*}
\frac{1}{\lb p_{n-1}}\int \limits_{0}^{a_{n-2}/\lb p_{n-2}} a_{n-1} d x_{n-2} &= \frac{1}{\lb p_{n-1}}\int\limits_{0}^{a_{n-2}/\lb p_{n-2}} (a_{n-2} - x_{n-2}\lb p_{n-2}) d x_{n-2}=\\
&= \frac{1}{\lb p_{n-1}}\frac{1}{2}\frac{a_{n-2}^2}{\lb p_{n-2}}.
\end{align*}
In general:
\begin{align*}
\int\limits_{0}^{a_{n-k}/\lb p_{n-k}} (a_{n-(k-1)})^{k-1} d x_{n-k} &= \int\limits_{0}^{a_{n-k}/\lb p_{n-k}} (a_{n-k} - x_{n-k}\lb p_{n-k})^{k-1} d x_{n-k}=\\
&=\frac{1}{\lb p_{n-k}}\frac{(a_{n-k})^k}{k}.
\end{align*}
Applying this formula until $k=n-1$ we get:
$$
C_{AHR}\approx \frac{(a_1)^{n-1}}{(n-1)\lb p_1}\cdots\frac{1}{3\lb{p_{n-3}}}\frac{1}{2\lb{p_{n-2}}}\frac{1}{\lb{p_{n-1}}}.
$$
Recalling that $a_1=\lb D$ we obtain:
$$
C_{AHR}\approx \frac{1}{(n-1)!}\prod\limits_{i=1}^{n-1}\frac{\lb D}{\lb{p_i}}.
$$

Let us perform somewhat more rigorous calculations. Let us consider the following sum:
\begin{align}
&\sum\limits_{x_{n-k}=0}^{\lfloor a_{n-k}/\lb p_{n-k}\rfloor} (a_{n-(k-1)})^{k-1} = \sum\limits_{x_{n-k}=0}^{\lfloor a_{n-k}/\lb p_{n-k}\rfloor} (a_{n-k} - x_{n-k}\lb p_{n-k})^{k-1} =\nonumber\\
&=\sum\limits_{x_{n-k}=0}^{\lfloor a_{n-k}/\lb p_{n-k}\rfloor}\sum\limits_{l=0}^{k-1} \binom{k-1}{l}(a_{n-k})^{k-1-l} (-1)^l x_{n-k}^l(\lb p_{n-k})^l\label{apdx:sum_1}
\end{align}
here $\binom{n}{k} = \frac{n!}{k!(n-k)!}$ are binomial coefficients. We use the following approximation of Faulhaber's formula (keeping only the leading power term):
\begin{align*}
\sum\limits_{x_{n-k}=0}^{\lfloor a_{n-k}/\lb p_{n-k}\rfloor} x_{n-k}^l &= \frac{1}{l+1}\sum\limits_{x_{n-k}=0}^{l}\binom{l+1}{x_{n-k}}B_{x_{n-k}}\left\lfloor \frac{a_{n-k}}{\lb p_{n-k}}\right\rfloor^{l-x_{n-k}+1}\approx\\
&\approx\frac{1}{l+1}\left\lfloor \frac{a_{n-k}}{\lb p_{n-k}}\right\rfloor^{l+1},
\end{align*}
here $B_{x_{n-k}}$ are Bernoulli numbers with the convention that $B_1 = +1/2$. As a result~\cref{apdx:sum_1} becomes:
\begin{align}
&\sum\limits_{x_{n-k}=0}^{\lfloor a_{n-k}/\lb p_{n-k}\rfloor} (a_{n-(k-1)})^{k-1} \approx\nonumber\\
&\approx\sum\limits_{l=0}^{k-1} \binom{k-1}{l}(a_{n-k})^{k-1-l} (-1)^l \frac{1}{l+1}\left\lfloor \frac{a_{n-k}}{\lb p_{n-k}}\right\rfloor^{l+1}(\lb p_{n-k})^l \approx\nonumber\\
&\approx \sum\limits_{l=0}^{k-1} \binom{k-1}{l}(a_{n-k})^{k-1-l} (-1)^l \frac{1}{l+1}\left( \frac{a_{n-k}}{\lb p_{n-k}}\right)^{l+1}(\lb p_{n-k})^l=\nonumber\\
&= \frac{(a_{n-k})^k}{\lb p_{n-k}}\frac{(-1)}{k}\sum\limits_{l=0}^{k-1} \frac{k(k-1)!}{l!(l+1)(k-(l+1))!}(-1)^{l+1}=\nonumber\\
&=\frac{(a_{n-k})^k}{\lb p_{n-k}}\frac{(-1)}{k}\sum\limits_{m=1}^{k} \frac{k!}{m!(k-m)!}(-1)^m=\nonumber\\
&=\frac{(a_{n-k})^k}{\lb p_{n-k}}\frac{(-1)}{k}(0-1) = \frac{(a_{n-k})^k}{\lb p_{n-k}}\frac{1}{k}.\label{apdx:sum_3}
\end{align}
Here we introduced $m=l+1$ and on the last step of derivation we used the following property of binomial coefficients:
$$
\sum\limits_{m=0}^{k} \binom{k}{m}(-1)^m= 0=1+\sum\limits_{m=1}^{k} \binom{k}{m}(-1)^m.
$$
As a result from~\cref{apdx:sum_3} we have the following approximate formula:
\begin{equation}
\sum\limits_{x_{n-k}=0}^{\lfloor a_{n-k}/\lb p_{n-k}\rfloor} (a_{n-(k-1)})^{k-1} \approx \frac{(a_{n-k})^k}{\lb p_{n-k}}\frac{1}{k}.\label{apdx:sum_4}
\end{equation}
Applying it sequentially for $k=1,\ldots,n-1$ we get the same approximation of~\cref{apdx:start}:
\begin{equation}
C_{AHR}\approx \frac{1}{(n-1)!}\prod\limits_{i=1}^{n-1}\frac{\lb D}{\lb{p_i}}.
\end{equation}

\section{\label{apdx:code}Description of the code of different algorithms implementation}
The program {\tt optimal\_size} is written in C programming language and used for generation of results in~\cref{sec:experiments} is provided in supplemental materials. It can be computed in GNU/Linux operation system using a command:
\begin{verbatim}
gcc -Ofast -flto optimal_size.c -o optimal_size -lglpk -lm
\end{verbatim}
The program requires GLPK to be installed in your system in some standard paths for library access.

These are the options used by author for compilation of the executable which generated results in~\cref{sec:experiments}. Program was linked with version 5.0 of GLPK, gcc had version 13.1.1. If the program is invoked without any command line parameters, short help is printed:
\begin{verbatim}
$ ./optimal_size 
Usage is the following:
./optimal_size target_positive_real_number prime_factor1 [prime_factor2 ... ]
\end{verbatim}
Here is an output for specific problem formulated in~\cref{sec:intro}:
\begin{verbatim}
$ ./optimal_size 478607.27 11 7 5 3 2
We are trying to compute optimal nonnegative integer powers for factors expansion:
(11)^{x_1}*(7)^{x_2}*(5)^{x_3}*(3)^{x_4}*(2)^{x_5} <= 478607.270000
Brute force algorithm with standard pow()-function.
Time for standard pow()-function brute force algorithm is 6.45620002300e-03 seconds.
(11)^{0}*(7)^{2}*(5)^{1}*(3)^{5}*(2)^{3} = 476280 is the closest choice for 478607.270000
Brute force algorithm with old integer pow()-function.
Time for old integer pow()-function brute force algorithm is 5.14299866300e-03 seconds.
(11)^{0}*(7)^{2}*(5)^{1}*(3)^{5}*(2)^{3} = 476280 is the closest choice for 478607.270000
Brute force algorithm with new integer pow()-function.
Time for new integer pow()-function brute force algorithm is 4.77261210300e-03 seconds.
(11)^{0}*(7)^{2}*(5)^{1}*(3)^{5}*(2)^{3} = 476280 is the closest choice for 478607.270000
Brute force algorithm.
Time for log2-based brute force algorithm is 2.75967481900e-03 seconds.
(11)^{0}*(7)^{2}*(5)^{1}*(3)^{5}*(2)^{3} = 476280 is the closest choice for 478607.270000
Recursive algorithm.
Time for recursive algorithm is 2.77179600000e-06 seconds.
(11)^{0}*(7)^{2}*(5)^{1}*(3)^{5}*(2)^{3} = 476280 is the closest choice for 478607.270000
Continuous LP algorithm.
Time for continuous LP algorithm using GLPK is 5.15028300000e-06 seconds.
(11)^{5}*(7)^{1}*(5)^{0}*(3)^{0}*(2)^{0} = 1127357 is the closest choice for 478607.270000
Integer LP algorithm.
Time for integer LP algorithm using GLPK is 7.08178717000e-04 seconds.
(11)^{0}*(7)^{2}*(5)^{1}*(3)^{5}*(2)^{3} = 476280 is the closest choice for 478607.270000
\end{verbatim}
Pay attention, this is just one run, performance was not necessarily optimal, like the numbers given in~\cref{tab:numerics}, which were chosen after several runs. Here is one extra result, specifically time for finding a solution through continuous linear programming solver and rounding all unknowns to the nearest integer, marked as {\tt Continuous LP algorithm}. As one can see result is ridiculous, even higher than original target number $D=478607.27$. The code can be configured to truncate (round down) the unknowns (this can be configures in the code through defining {\tt TRUNC\_LP} preprocessor option in the beginning of the code), which gives as ridiculous result as in previous case. This computation was added as a mere demonstration that continuous formulation with rounding or truncation has no sens for this particular problem. Few other preprocessor options are well commented in code.

\section*{Acknowledgments}
This paper was stimulated by previous papers~\cite{2DFourier2020,KLSD2023}, which originated in research partially funded by Simons' Foundation through Simons' Collaboration on Wave Turbulence (award \#651459). Author would like to thank Prof. Evangelos Coutsias from Stony Brook University, SUNY, NY, for enlightening discussions on linear programming.


\end{document}